\documentclass{article}
\usepackage{graphicx}
\usepackage{cite}
\usepackage{epsfig}
\usepackage{amsfonts}
 \usepackage{amssymb}
\usepackage{amsmath}
\usepackage{slashed}

\usepackage{hyperref}
\usepackage{graphicx}       
\usepackage{dcolumn}        
\usepackage{bm}            		 
\usepackage{amssymb}
\usepackage{amstext}
\usepackage{tensor}
\usepackage{color} 
\usepackage{transparent}

\date{\today}

\begin{document}
\begin{center}

{\Large \bf Spontaneous vectorization of electrically charged black holes}
\vspace{0.8cm}
\\
{Jo\~ao M. S. Oliveira$^{\dagger}$, 
Alexandre M. Pombo$^{\ddagger}$, \\
\vspace{0.3cm}
$^{\dagger}${\small Centro de Astrof\'\i sica e Gravita\c c\~ao - CENTRA,} \\ {\small Departamento de F\'\i sica,
Instituto Superior T\'ecnico - IST, Universidade de Lisboa - UL,} \\ {\small Avenida
Rovisco Pais 1, 1049-001 Lisboa, Portugal}
\vspace{0.3cm}
\\
$^{\ddagger }${\small Departamento de Matem\'atica da Universidade de Aveiro and } \\ {\small  Centre for Research and Development  in Mathematics and Applications (CIDMA),} \\ {\small    Campus de Santiago, 3810-183 Aveiro, Portugal}
}
\end{center}

	\begin{abstract}   
    In this work, we generalise the spontaneous scalarization phenomena in Einstein-Maxwell-Scalar models to a higher spin field. The result is an Einstein-Maxwell-Vector model wherein a vector field is non-minimally coupled to the Maxwell invariant by an exponential coupling function. We show that the latter guarantees the circumvention of an associated no-hair theorem when the vector field has the form of an electric field. Different than its scalar counterpart, the new \textit{spontaneously vectorized} Reissner-Nordstr\"om (RN) black holes are, always, undercharged while being entropically preferable. The solution profile and domain of existence are presented and analysed.
	\end{abstract}

\medskip
%
\section{Introduction}
%
	The recent developments in black hole (BH) detection through gravitational wave emission and imaging led to one of the most important observational advances in BH history. Never before was it possible to study highly compact objects with such precision and certainty, allowing the possibility to discern between alternative theories of gravity through BH observations. One of the key differences between these theories is the phenomena of spontaneous scalarization/vectorization, which can be contained in the more generic spontaneous tensorization phenomena.    

	Spontaneous scalarization has been thoroughly discussed in the literature, with the most recent development in extended Scalar-Tensor-Gauss-Bonnet (eSTGB) gravity \cite{Doneva:2017bvd,Silva:2017uqg,Antoniou:2017acq}. In the latter, the model contains a scalar field non-minimally coupled to the Gauss-Bonnet gravity correction term. While this scalarization phenomenon is induced by gravity, the original spontaneous scalarization mechanism \cite{Damour:1993hw} was induced through the presence of matter.
	
	Alternatively, one can non-minimally couple the scalar field to the Maxwell invariant introducing the Einstein-Maxwell-Scalar (EMS) model~\cite{Herdeiro:2018wub}. The mechanism was also considered for Kerr BH where scalarization can be spin-induced ~\cite{Cunha:2019dwb,Herdeiro:2020wei,Berti:2020kgk,dima2020spin}. In all cases, the phenomena occur due to a tachyonic instability that emerges for certain BH configurations. The tachyonic instability induces an exponential growth of the scalar field, and the resulting BH solution is now immersed in a scalar cloud.

	While the eSTGB and Kerr model seems to be more astrophysically relevant in comparison to the EMS models, the latter is much simpler to deal with, while still presenting the same phenomenological behaviours. The properties of the various types of scalarized EMS BHs have been extensively studied \cite{fernandes2019spontaneous,astefanesei2019einstein,fernandes2019charged,blazquez2020einstein,blazquez2020quasinormal,Astefanesei:2020qxk}.
	
 When considering further generalisations of this mechanism, it is natural to wonder if the spontaneously growing matter can be a vector or even a tensor. In this paper, we consider the phenomenon of \emph{spontaneous vectorization (SV)}. 

	Vector fields and their role in extended theories of gravity have been discussed before~\cite{baum1970vector, gleiser2005linear}, and examples of BHs with vector hair have also been found~\cite{herdeiro2016kerr,fan2016black}. The phenomenon of vectorization was later considered in the extended Vector-Tensor-Gauss-Bonnet (eVTGB) theory~\cite{annulli2019electromagnetism,ramazanouglu2017spontaneous,Ramazanoglu:2019gbz} (the vector analogue to the eSTGB theory); in theories non-minimally coupled to matter~\cite{minamitsuji2020spontaneous,Kase:2020yhw};
 		 and other theories of gravity~\cite{annulli2019electromagnetism,Ramazanoglu:2019jrr}. The main idea of this paper is to consider the EMS mechanism used in~\cite{Herdeiro:2018wub} and extend it to an \textit{Einstein-Maxwell-Vector (EMV)} model. For that let us consider the action:
	\begin{equation}\label{action}
	\mathcal{S} = \frac{1}{4\pi}\int d^4 x \sqrt{-g}\bigg[\frac{R}{4} - \frac{f(|B|^2)}{4}F^{\mu\nu}F_{\mu\nu} - \frac{1}{4}G^{\mu\nu}G^*_{\mu\nu} 
	\bigg] \ ,
	\end{equation}
	 where $R$ is the Ricci tensor, the Maxwell field strength $F_{\mu\nu} = \partial_\mu A_\nu-\partial_\nu A_\mu$, and $G_{\mu\nu}= \partial_\mu B_\nu-\partial_\nu B_\mu$ which represents the field strength of a (possibly complex) vector field $B_\mu$. While $A_\mu$ and $B_\mu$ are both vector fields, for nomenclature simplicity, from now on, we will refer to $A_\mu$ as Maxwell field and $B_\mu$ as vector field. 
	 
	The vector field is non-minimally coupled to the Maxwell term through the coupling function $f(|B|^2)$, where $|B|^2 = B ^\mu B^*_\mu$, and
	\begin{equation}
	f(0)=1 	\ ,
	\end{equation}
	so that we may recover Einstein-Maxwell when we have a trivial 
field $B^\mu$. Note that this is a straightforward generalisation of the massless scalar case, where a coupling $f(\phi)$ is considered.

	This paper is organized as follows: Sec.~\ref{S2} is dedicated to an analytical study of the model \eqref{action}. A no-vector-hair theorem for a spherically symmetric ansatz is presented and followed by a small discussion of the flat spacetime case in Sec.~\ref{S3}. The existence line is presented in Sec.~\ref{S4} as well as the full non-linear model and quantities of interest. The numerical solutions are presented and studied in Sec.~\ref{S5} and we conclude in Sec.~\ref{S6}.
%
\section{The model}\label{S2}
%
	The stress-energy tensor for the model described by the action~\eqref{action} is:
	\begin{align}\label{stress}
	T_{\mu\nu} =& f(|B|^2)\bigg(F_\mu^{\,\,\alpha}F_{\nu\alpha} - \frac{1}{4}g_{\mu\nu}F^{\alpha\beta}F_{\alpha\beta}\bigg) + \frac{1}{2}\bigg(G_\mu^{\,\,\alpha}G^*_{\nu\alpha}+G^{*\alpha}_\mu G_{\nu\alpha}-\frac{1}{2}g_{\mu\nu}G^{\mu\nu}G^*_{\mu\nu}\bigg) \nonumber \\
	 &+ \frac{1}{4}\frac{df}{d|B|^2}F^{\alpha\beta}F_{\alpha\beta}\big(B_\mu B^*_\nu+ B^*_\mu B_\nu) 
	 \ .
	\end{align}
	Note that, compared to the scalar case, the need to consider the scalar $g^{\mu\nu}B_\mu B^*_\nu$ in the coupling $f(|B|^2)$ introduces the last term in \eqref{stress}. The fact that it can be negative allows the possibility of a violation of the weak energy condition.

	The massless vector field $B_\mu$ equations, which is described by a (massless) Proca equation, and the Einstein equations come as:
	\begin{align}\label{EProca}
	\nabla _\mu G^{\mu\nu} &= \frac{1}{2}\frac{df}{d|B^2|}F^2
	B^\nu \ , \\
	R_{\mu\nu}&-\frac{1}{2}g_{\mu\nu}R = 2T_{\mu\nu} \label{EFEquations}\ .
	\end{align}
	The vector field $B^\mu$, while being massless, due to the interaction with the electromagnetic field, gains an effective mass $(\mu _{eff})$
	\begin{equation}
	\mu^2_{eff} = \frac{1}{2}\frac{df}{d|B|^2}F^2
	 \ ,
	\end{equation}
	which, for certain forms of the coupling function and the electromagnetic field, can be negative. This translates into a \emph{tachyonic instability}, \textit{i.e.,} for an initial trivial configuration of $B_\mu$, corresponding to a Reissner-Nordstr\"om (RN) spacetime, a small vector field perturbation, $\delta B_\mu$, grows exponentially and drives the system away from the RN solution. The result is a \textit{spontaneously vectorized RN BH} (VRN).

	For a purely electric configuration $F^2<0$: $\mu _{eff} ^2 <0$ requires
	\begin{equation}
	\frac{df}{d|B|^2}>0 \ ,
	\end{equation}
	and the opposite sign for a purely magnetic configuration.	For a deeper study on the several possible coupling function solutions in the scalarized case see \cite{astefanesei2019einstein} (the same line of thought can be applied here). Both conditions are satisfied by a quadratic exponential coupling
	\begin{equation}
	f(|B|^2) = e^{\alpha |B|^2} \ .
	\end{equation}
	For this coupling, spontaneous vectorization of a purely electric RN BH occurs for $\alpha>0$.
	
	Another important property of the model is that the Lorenz condition is not implied by the equations. of motion. If we take the divergence of \eqref{EProca}
	\begin{equation}
	 \nabla^\mu B_\mu = - \frac{\nabla^\mu (\mu^2_{eff})}{\mu^2_{eff}}B_\mu \ ,
	\end{equation}	
	which, as we can see, does not correspond to the Lorentz condition, since $\mu^2_{eff}$ is now a function.

	\medskip
	
	The metric ansatz of a static, spherically symmetric spacetime can be described by
	\begin{equation}\label{mansatz}
	ds^2 = - \sigma(r)^2N(r) dt^2 + \frac{dr^2}{N(r)} + r^2\big( d\theta^2+\sin^2\theta d\phi^2\big) \ , \qquad \qquad \mathrm{with} \qquad N(r)=1-\frac{2m(r)}{r} ,
	\end{equation}
	where $m(r)$ is the Misner-Sharp mass function~\cite{misner1964relativistic}, $r$ is the areal radius and $\sigma (r)$ is a real metric function.

	The vector field inherits the metric spherical symmetry and, therefore, for its ansatz we consider
	\begin{equation}\label{Bansatz}
	B(r,t) = \big[ B_t(r)dt + i B_r(r)dr\big] e^{-i\omega t} \ ,
	\end{equation}
	while the Maxwell field will only have an electrical component
	\begin{equation}\label{Aansatz}
	A(r) = A_t (r) dt \ ,
	\end{equation}

%
\section{Vector theorems}\label{S3}
%

%
	\subsection{No-vector-hair theorem}\label{S31}
%
	For this section, let us follow the work done by Herdeiro \textit{et al.}~\cite{herdeiro2016kerr} and generalise their results for our current model.\\
	\textbf{Theorem:} \textit{A spherically symmetric, static, asymptotically flat and electrically charged BH spacetime, regular on and outside the event horizon, which solves the Einstein-Maxwell complex-Proca field equations, and for which the massive Proca field inherits the spacetime spatial symmetries but can have an harmonic time dependence of the type $e^{-i\omega t}$ with $\omega\neq 0$, cannot support a non-trivial, finite on and outside the horizon, Proca field.}
	
	To prove this argument, we will follow~\cite{Pena:1997cy}. If the vector field is given by \eqref{Bansatz}, the Proca equations. in the metric \eqref{mansatz} are
	\begin{align}
	\frac{d}{dr}\left[\frac{r^2\Big( B_t'(r)-\omega B_r(r)\Big) }{\sigma(r)}\right] = \frac{\mu_{eff}^2 r^2 B_t(r)}{\sigma(r) N(r)} \label{Peq1}\ ,\\
	B_t'(r) = \omega B_r(r)\left(1-\frac{\mu_{eff}^2\sigma^2(r)N(r)}{\omega^2}\right) \label{Peq2}\ .
	\end{align}

	For a BH solution, we assume the existence of an outermost horizon at $r=r_H>0$, which requires $N(r_H)=0$. Every $r>r_H$ surface will then be a timelike surface and $N'(r_H)>0$. As the sign of $\sigma$ is irrelevant to the equations of motion, we can, without loss of generality, consider $\sigma(r_H)>0$.
	
	 The proof comes as follows: consider that there is a small enough region close to the horizon, $r_H< r < r_1$, for which $\mu^2_{eff}<0$ (which is always verified for a massless field). This means that
	\begin{equation}\label{condp}
	1-\frac{\mu_{eff}^2\sigma^2(r)N(r)}{\omega^2}>0 \ ,
	\end{equation}
	is guaranteed in this region. Equation \eqref{Peq2} then implies that the sign of $B'_t$ is equal to the sign of $B_r$. If we integrate Eq. \eqref{Peq1} in an interval $[r,r_c]	\subset\ ]r_H,r_1[$ and replace $B_t'(r)$ by Eq.~\eqref{Peq2}, we get
	\begin{equation}\label{condp2}
	B_r(r) = -\frac{\omega}{r^2 \mu_{eff}^2\sigma(r)}\int_r^{r_c}dr\frac{\mu_{eff}^2 r^2 B_t(r)}{\sigma(r) N(r)} \ ,
	\end{equation}
	which imposes that the sign of $B_r$ must be opposite to the sign of $B_t(r)$.
	
	The theorem is now proven by contradiction. As we will see ahead, $B_t$ must be zero at the horizon. So, if $B_t'>0$ close to the horizon, then $B_t>0$ in this region. However, as we know from the considerations above, $B_t'$ has the same sign as $B_r$ implying that $B_r>0$ which is the same sign as $B_t$, contradicting the equation above. The exact same reasoning applies if we consider $B_t'<0$, meaning that the only BH solution compatible with the conditions above is when $B_t=0=B_r$: the Reissner-Nordstr\"om family of solutions.
	
	This same theorem can be generalised for the case where $\mu^2_{eff}>0$ (for example, if we have a massive field\footnote{If the $B$ vector was massive, $\mu^2_{eff}$ would instead take the form $\mu^2_{eff}= \mu_B^2 +\frac{1}{2}\frac{df}{d|B|^2}F^2$, and $\alpha _{min}$ would be $\mu_B$ dependent.}) in the region $r_h<r<r_1$. As long as this region is small enough, we can always satisfy condition~\eqref{condp}. The fact that $N(r_H)=0$, implies that the \textit{l.h.s.} of condition~\eqref{condp} is very close to unity in this region. Since Eq.~\eqref{condp2} is independent of the sign of $\mu^2_{eff}$, the rest of the theorem follows.
	
	Note that, this theorem is not valid for $\omega=0$. The latter imposes a solely $r$ dependent vector field. In that case, we can obtain the equation for $B_r$ from the Proca Eq.~\eqref{EProca}
	\begin{equation}
	\nabla_t G^{tr} = 0 = \mu^2_{eff}B^r \ .
	\end{equation}
	Since $\mu^2_{eff}$ is assumed to be non-zero, we have that the radial component $B^r$ must vanish.
	Then, the only viable vector field ansatz is
	\begin{equation}\label{Btansatz}
	B(r) = B_t(r)dt \ .
	\end{equation}
%

%
	\subsection{Flat spacetime electric no go theorem}\label{S32}
%
	Let us now consider the real ansatz \eqref{Btansatz} for the vector field. By assuming a purely electric field, given by \eqref{Aansatz}, the electromagnetic equation of motion is
	\begin{equation}
	\nabla_\mu(f F^{\mu\nu}) = 0 \Rightarrow A_t' = \frac{Q}{r^2 f} \ .
	\end{equation}
	The Virial identity on flat spacetime is
	\begin{equation}\label{VIdFlat}
	\int_{0}^\infty dr\frac{1}{r^2} \left(r^4 B_t'^2+\frac{Q^2}{f}\right) = 0 \ .
	\end{equation}
	Since both terms are always positive (for $f>0$), we find that the Virial identity can only be respected for the trivial configuration $B'_t=0$ and $Q=0$. When $Q=0$, the effective mass term of $B$ vanishes, so $B$ gains gauge freedom and becomes a typical Maxwell field, allowing us to set $B_t =0$.
	
	Alternatively, we can see this through a map to a scalar field, $\Phi$. If we consider the effective action for this configuration
	\begin{equation}
	\mathcal{S} = \frac{1}{4\pi}\int d^4 x \bigg[- \frac{1}{2}\frac{Q^2}{fr^4} + \frac{1}{2}(\partial_r B_t)^2%
	\bigg] \ ,
	\end{equation}
	and the mapping $B_t(r)\rightarrow i\Phi(r)$, one recovers the effective action for the static, spherically symmetric EMS model in flat spacetime
	\begin{equation}
	\mathcal{S} = \frac{1}{4\pi}\int d^4 x \bigg[- \frac{1}{2}\frac{Q^2}{fr^4} - \frac{1}{2}(\partial_r \Phi)^2%
	\bigg] \ .
	\end{equation}
	This means that the $B$ field with the ansatz \eqref{Btansatz} acts as a \emph{ghost} scalar field. Flat, spherically symmetric spacetime solutions with a real scalar field $\Phi(r)$ have been found for this model in \cite{Herdeiro:2019iwl} for an arbitrary coupling $f(\Phi)$. These solutions are then mapped to purely imaginary $B_t(r)$ solutions which do respect the Virial identity \eqref{VIdFlat}.

%
\section{Spontaneous Vectorization}\label{S4}
%
%
	\subsection{Bifurcation points}\label{S41}
%
	In the absence of backreaction, the EMV model can be seen as a Reissner-Nordstr\"om BH that suffers a perturbation from a vector field $B_\mu$. In this case, the line element is the same as the RN BH
	\begin{equation}
	 ds^2 = -N(r) dt^2+\frac{dr ^2}{N(r)}+r^2 \big( d\theta^2+\sin^2\theta d\phi^2\big)\ ,\qquad \mathrm{with}\qquad N(r)=1-\frac{2M}{r}+\frac{Q^2}{r^2}\ , 
	\end{equation}
	where $M$ ($Q$) is the ADM mass (electric charge) of a RN BH. In this study we will consider the full model \eqref{action}, however the coupling function needs to be linearly approximated in $|B|^2$ as $f(|B|^2)= e^{\alpha |B|^2} \approx 1+\alpha |B|^2$.

	The Proca Eq.~\eqref{EProca} that describes a nodeless, massless vector field coupled to the Maxwell invariant and has the form \eqref{Btansatz}, comes as:
	\begin{equation}\label{EProca2}
	\frac{g^{rr}}{\sqrt{-g}}\partial _r \Big[ \sqrt{-g}\partial ^r B_t(r)\Big]+\alpha \frac{Q^2}{r^4} B_t(r) =0\ ,\qquad \qquad \mathrm{with} \qquad \mu _{eff} ^2 = -\alpha \frac{Q^2}{r^4}\ .
	\end{equation}
	A RN solution that supports SV requires an effective mass $\mu _{eff} ^2<0$, and field equation reduces to an eigenvalue problem in $M$
	\begin{equation}\label{EProca3}
	r^2 B_t ''+2 r B_t ' +\frac{\alpha  Q^2 }{r (r-2 M)+Q^2} B_t = 0\ .
	\end{equation}
	While the equations for $B^t$ is easier to deal with, due to the divergence at the horizon of $g^{tt}$ the value of $B^t (r_H)$ is not well defined. At the horizon, the physical vector field obeys $B_t=g_{tt}B^t(r_H)=0$ as well as at infinity $B_t (r\rightarrow \infty) =0$. Close to the horizon, the vector field can be approximated as
	\begin{equation}\label{EExist}
	B_t (r) \approx b_1 (r-r_H) -b_1 \frac{ r_H (\alpha -2) Q^2}{2 \left(r_H ^2-Q^2 \right)} (r-r_H) ^2 + \cdots\ , \qquad \qquad \mathrm{with} \qquad M=\frac{Q^2+r_H ^2}{2 r_H}\ ,
	\end{equation}
	The field Eq.~\eqref{EProca3} has an analytical solution that obeys the proper boundary conditions Eq.~\eqref{EExist} 
	\begin{equation}
	B_t= z \, _2F_1\left[\frac{1}{4} \left(3-y\right);\frac{1}{4} \left(3+y\right);2;-z\right] \ ,
	\end{equation}	
	with
	\begin{equation}
	z = 4 Q^2 r_H \left(\frac{Q^2 r_H}{r^2}-\frac{Q^2}{r}-\frac{r_H ^2}{r}+r_H\right)\ ,\qquad \qquad \mathrm{and} \qquad y= \sqrt{4 \alpha-1}\ ,
	\end{equation}
	and $_2 F _1$ is an hypergeometric function. Observe that $\alpha _{min} \geqslant \frac{1}{4}$, which occurs for an extremal RN configuration (first bifurcation point), while $Q\rightarrow 0$ implies $\alpha \rightarrow \infty$. Observe that for each value of $\alpha$ and $Q$, Eq.~\ref{EExist} yields a value of $M$ at wich the VRN solution bifurcates from the RN BH. The computation of all bifurcation points for a range of $\alpha$ gives then the \textit{existence line}.

%
	\subsection{The full non-linear model}\label{S42}
%
	The set  of full non-linear field equations that result from the model \eqref{action} with the ansatz \eqref{mansatz} and \eqref{Btansatz} are
	\begin{align}
	m' &= \frac{N r^4  B_t ^{' 2}-Q^2 e^{\frac{\alpha  B_t ^{2}}{N \sigma ^2}} \left(2 \alpha  B_t^ {2}-N \sigma ^2\right)}{2 N r^2 \sigma ^2} \ ,\qquad \qquad \qquad \qquad \sigma ' = -\frac{B_t ^2 \alpha  Q^2 e^{\frac{\alpha  B_t ^2}{N \sigma ^2}}}{r^3 N^2 \sigma }\ ,\nonumber\\
	A_t ' &= -\frac{Q \sigma e^{\frac{\alpha  B_t ^2}{N \sigma ^2}}}{r^2}\ ,\qquad \qquad \qquad \qquad\qquad \qquad \qquad \qquad \qquad B_t'' = B_t ' \left(\frac{\sigma '}{\sigma }-\frac{2}{r}\right)- \frac{\alpha  Q^2 e^{\frac{\alpha  B_t ^2}{N \sigma ^2}}}{r^4 N}B_t \ .
	\end{align}
	where the Maxwell potential $A_t'(r)$ is under a first integral that was used to simplify the other field equations. For notation simplicity, we let the radial dependence fall. Close to the horizon, the metric functions and vector field can be approximated by a power series as
	\begin{eqnarray}
	m &\approx & \frac{r_H}{2}+\frac{\frac{b_1 ^2 r_H^4}{\sigma _0 ^2}+Q^2}{2 r_H ^2} (r-r_H)+\cdots\ ,\ \qquad \qquad \sigma \approx \sigma _0- \frac{b_1 ^2 r_H ^3 \sigma _0 ^3 \alpha  Q^2}{\big(b_1 ^2 r_H ^4+\sigma _0 ^2 (Q^2-r_H ^2)\big)^2}(r-r_H)+\cdots\ ,\nonumber \\
	A_t &\approx & -\frac{Q\sigma _0}{r_H^2} (r-r_H) +\cdots \ , \qquad \quad \qquad \qquad \qquad B_t  \approx  b_1 (r-r_H)+ b_2 (r-r_H)^2 +\cdots\ ,\nonumber\\ 
	b_2 &=& b_1 \left[ \frac{\alpha  Q^2 \sigma _0 ^4 (Q^2-r_H ^2)}{2 r_H \big( b_1 ^2 r_H ^4+\sigma _0 ^2 (Q^2-r_H ^2)\big)^2}-\frac{1}{r_H}\right]\ ,
	\end{eqnarray}
	with $b_1$ the value of the vector field derivative and $\sigma _0$ the value of the $\sigma$ function, at the horizon. At infinity, we impose asymptotical flatness and the metric/field functions can be approximated by 
	\begin{eqnarray}
	 m &\approx & M+\frac{Q^2+P^2}{2r}+\cdots\ ,\qquad \qquad \qquad \sigma \approx 1- \frac{Q^2\alpha }{2 r^2}\cdots\ ,\nonumber \\
	 A_t  &\approx & \Phi _e -\frac{Q}{r}+\cdots \ ,\qquad \qquad \quad \qquad \qquad B_t \approx \frac{P}{r} \ ,
	\end{eqnarray}
	with $\Phi _e$ the electrostatic potential difference at infinity and $P$ the ``vector charge`` obtained from the asymptotic decay. 
	
%
	\subsection{Quantities of interest and Smarr law}\label{S43}
%
	Two horizon quantities of interest are the Hawking temperature and the horizon area
	\begin{equation}
	T_H = \frac{1}{4\pi} N'(r_H) \sigma _0\ , \qquad \qquad A_H = 4\pi r_H ^2\ ,
	\end{equation}
	these, together with the horizon vector field derivative, $b_1$, and the horizon $\sigma$ value, $\sigma _0$, describe the relevant horizon data.

	The variation of the ADM Mass is described by the first law: $dM=T_HdS+\Phi _e dQ$. The vectorized solutions obey the Smarr law
	\begin{equation}
	 M=\frac{1}{2}T_H S_H+ \Phi _e Q +M_P\ ,
	\end{equation}		
	where $M_P$ is the energy stored in the surrounding vector field, which can be computed through a Komar integral
	\begin{equation}
	 M_P = -\int _\Omega dr d\theta d\phi \sqrt{-g}\left( 2T_t ^t-T\right)
	 	 = -4\pi \int _{r_H} ^\infty dr \frac{Q^2  \left(\alpha  B_t ^2-N \sigma ^2\right)-r^4 N B_t^{'2}}{fN r^2 \sigma }\ ,
	\end{equation}
	with $T$ the trace of the stress-energy tensor.
	
	In addition, the solutions satisfy the Virial identity, wich is obtained by a Derrick-type~\cite{derrick1964comments} scaling argument, 
	\begin{equation}
	\int_{r_H}^\infty dr\left[(r-r_H)\frac{\alpha Q^2B_t^2}{2 r^3 N f \sigma} + (2r_H-r) N^2\left(r^4 B_t'^2+\frac{Q^2 \sigma^2}{f} \right)\right]=0 \ .
	\end{equation}
	The generic vectorized solution is not known in closed form, and a numerical approach is necessary. To solve the latter, we use an adaptative step $6(5)^{th}$ order Runge-Kutta integrator (local error of $10^{-20}$), while the boundary conditions are imposed through a Secant shooting strategy with a tolerance of $10^{-12}$ in terms of the unknown parameters $b_1 $ and $\sigma _0$. In all the presented solutions the Virial identity gave an error of $\sim 10^{-8}$, while the Smarr law gave $\sim 10 ^{-4}$.

	At last, observe that the model possesses the scaling symmetry	$r\rightarrow \lambda r$, $Q\rightarrow \lambda Q\,$ where $\lambda >0$ is a constant. Under this scaling symmetry, all other quantities change accordingly, \textit{e.g.}, $M\rightarrow \lambda M$, while the coupling function $f(|B|^2) $ remains unchanged. For the physical discussion let us introduce the reduced quantities
	\begin{equation}
	q\equiv \frac{Q}{M}\ , \qquad a_H \equiv \frac{A_H}{16\pi M^2}\ ,\qquad t_H \equiv 8\pi T_H M\ .
	\end{equation}
%
	\subsubsection*{Light Rings}
%
	One of the most important astrophysical properties of BHs is the presence of a \textit{light ring (LR)} -- since we are dealing with spherical symmetry, the LR is, in fact, a sphere: a photon sphere. To find the LR radius, $r_{LR}$, of a spherical spacetime, one must consider the null geodesics ($ds^2=0$) of the metric ansatz \eqref{mansatz} (the dot represents a derivative with respect to an affine parameter):
	\begin{equation}
	\dot{r}^2 = \frac{E^2}{\sigma^2} - \frac{l^2N}{r^2} \ ,
	\end{equation}
	where $E$ and $l$ represent the energy and angular momentum of a photon along the geodesic. The LR is circular, implying $\dot{r}=0$ and $\ddot{r}=0$. The first condition relates the energy with the angular momentum of the photon $E=l\sqrt{N}\sigma/ r$ while the second gives us the condition necessary to find $r_{LR}$:
	\begin{equation}
	\sigma\left(-2m'+\frac{2m}{r}\right)+2\left(1-\frac{2m}{r}\right)(r\sigma'-\sigma) = 0 \ .
	\end{equation}
	For the RN metric we have $\sigma=1$ and $m(r) = M -Q^2/2r$, giving us 
	\begin{equation}
	r^{RN}_{LR} = \frac{3M\pm\sqrt{9M^2-8Q^2}}{2} \ .
	\end{equation}
	As demonstrated in \cite{cunha2017light,cunha2020stationary}, LR always come in pairs. For a BH, one of the LR is inside and the other outside the external horizon. 
	
%
\section{Numerical results}\label{S5}
%

%
	\subsection{Solutions profile}\label{S51}
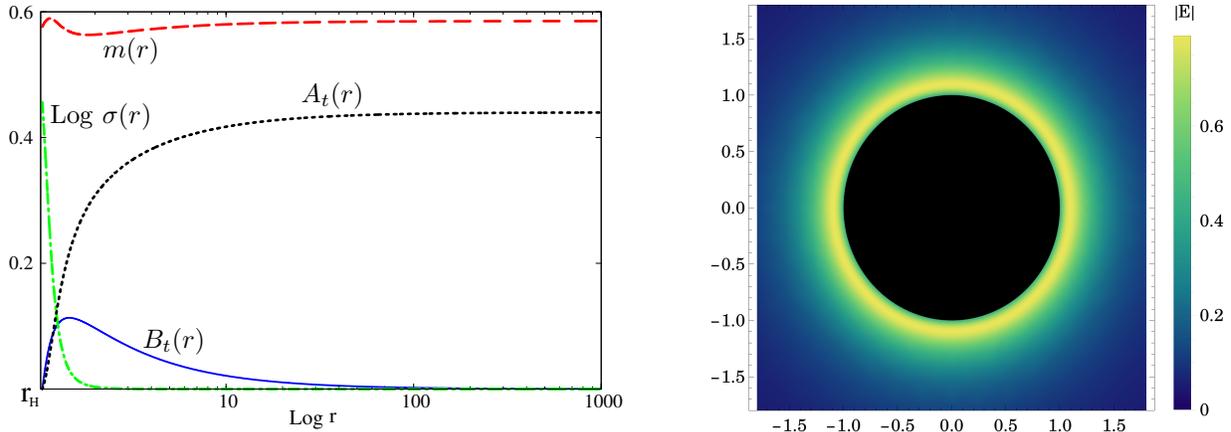
\begin{figure}[h]
	 \centering
	  \begin{picture}(0,0)
	   \put(45,145){$m(r)$}
	   \put(120,128){$A_t (r)$}
	   \put(60,35){$B_t (r)$}
	   \put(25,120){$\mathrm{Log}\ \sigma (r)$}
	  	\end{picture}
	 	 \includegraphics[scale=0.66]{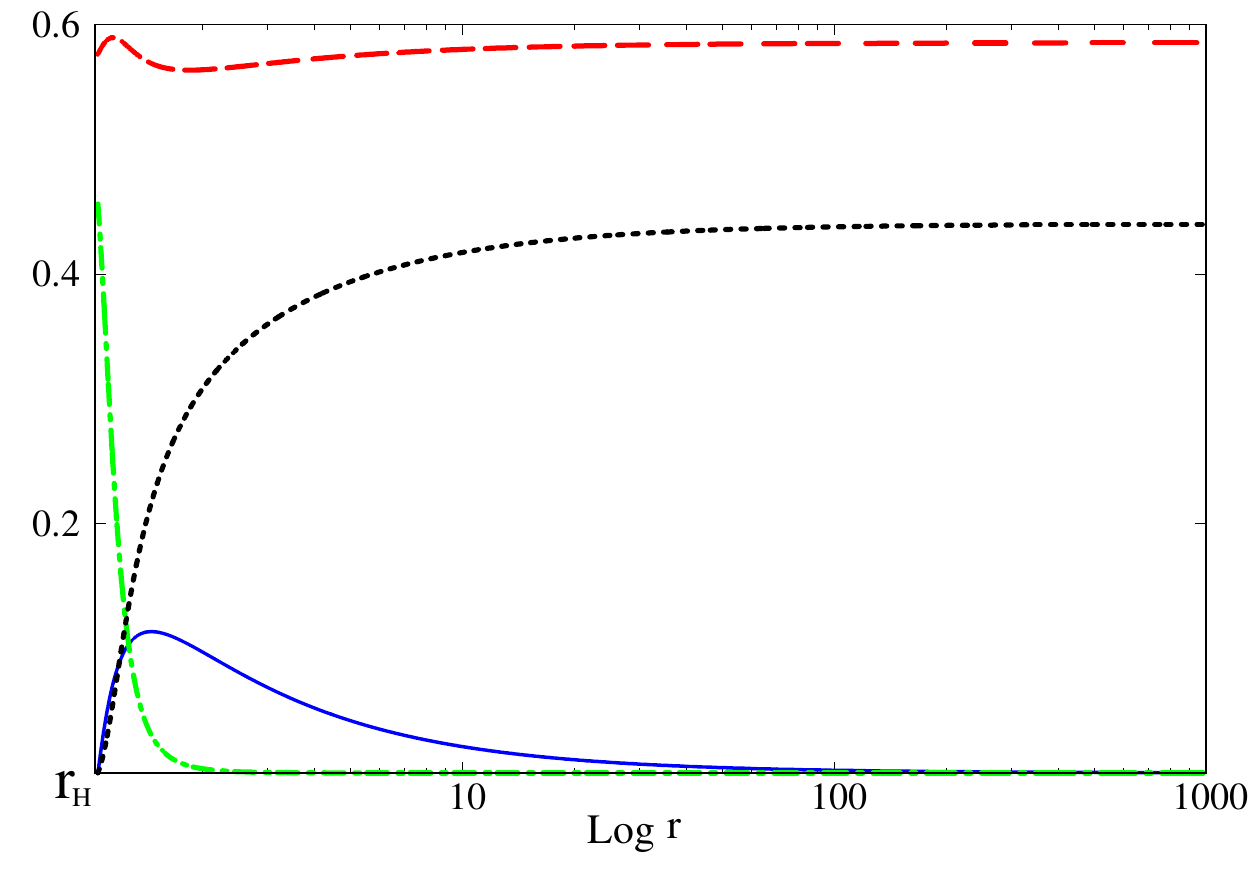}\hfill
	 	 \includegraphics[scale=0.115]{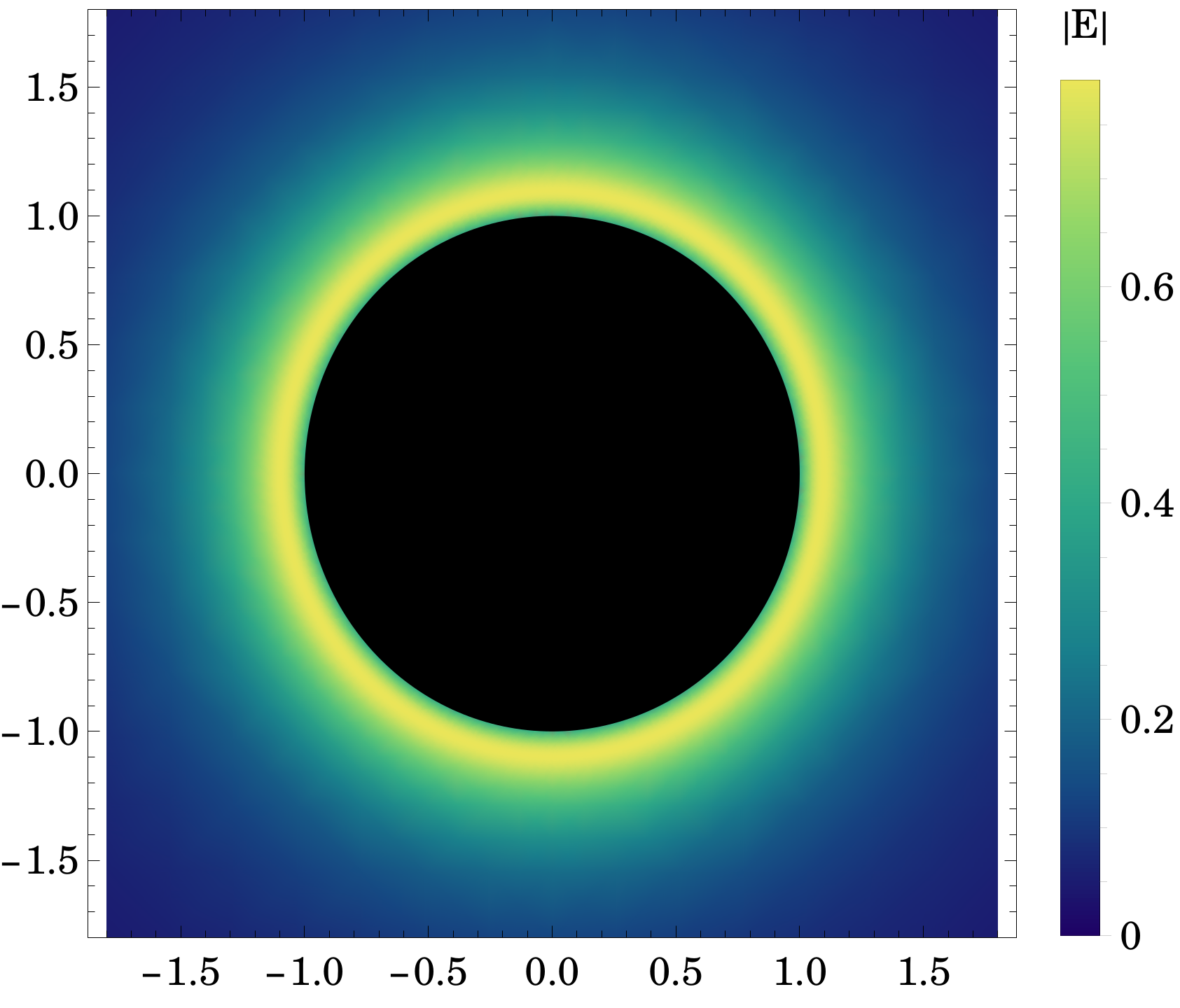}
	 \caption{(Left) graphical representation of the field functions profile and (right) density plot of the electric field strength along the equatorial plane for $\alpha =25$, $r_H=1.0$ and $Q=0.25$.}
	 \label{F2}
	\end{figure}

%
	Let us start by studying the generic behaviour of the metric functions and the vector fields of a fundamental (nodeless) state VRN BH. In Fig.~\ref{F2} is represented the radial dependence of the various field functions for an illustrative solution with $\alpha =25$, $Q=0.25$, $r_H=1.0$, and a charge to mass ratio $q=0.4271$.

	A universal feature of the fundamental solutions is the existence of a bulge of $B_t$ around the event horizon. Since regularity imposes a null vector field at $r_H$ and infinity, the only non-trivial, nodeless vector field solutions possess a sharp increase very close to the horizon ($b_1 >0$), reaches a maximum and then ``slowly`` decays as $r^{-1}$. 
	
	Numerical analysis shows that, for a fixed charge and mass, increasing $\alpha$ corresponds to an increase of the magnitude of the field $B_t$. The distance of the maximum of $B_t$ relative to the horizon is also observed to increase slightly along with an increase in $\alpha$.

	Besides, due to the coupling with the Maxwell field, there will be a modulation of the electric potential that ultimately creates a non-monotonic electric field that can have important implications in the accretion disk formation (see Fig.~\ref{F2} right). As a contrast, the electro-vacuum RN solution, as well as in the scalarized case, is monotonically crescent and the modulation associated with the latter is closer to a damping.

	One other interesting characteristic of this model is the presence of a region with negative energy density. Observe the $m(r)$ profile in Fig.~\ref{F2} right. In the latter, there is a valley, which corresponds to the region where the $B_t$ reaches its peak. This can be easily understood by observing that the Komar mass (mass associated with the external vector field), due to the negative energy density term in the EM Tensor \eqref{stress}, gives a negative contribution to the ADM mass, violating the weak energy condition.

	Regarding the light ring radii, we show some values in Tab.~\ref{T1}. We can see that there is an increase of the LR radius when we increase the coupling constant $\alpha$ and that the LR radii of vectorized BHs are smaller than the corresponding RN black holes.
\begin{table}[h]
\begin{center}
\caption{\label{T1}Light ring radii for four $\alpha $ values with $Q=0.25$ and $r_H=0.52$.}
\vspace{2mm}
\begin{tabular}{ c |c |c|c|c }
 	$\alpha$ & 6 & 8 & 10 & 12 \\ 
 	\hline
 	$r_{LR}$ & $0.73$ & $0.77$ & $0.79$  & $0.80$ \\ 
 	$r_{LR}/r^{RN}_{LR}$ & $0.77$ & $0.83$ & $0.84$ & $0.88$ 
\end{tabular}
\end{center}
\end{table}

%
	\subsection{Domain of existence}\label{S52}
%
	Generating several solutions allows us to obtain a region of the domain of existence for the VRN BH solutions. The latter is delimited by the existence line -- at which $b_1\rightarrow 0$ -- and a critical line -- with $b_1 \rightarrow \infty$.

	\begin{figure}[h]
	 \centering
	  \begin{picture}(0,0)
	   	\end{picture}
	 \includegraphics[scale=0.66]{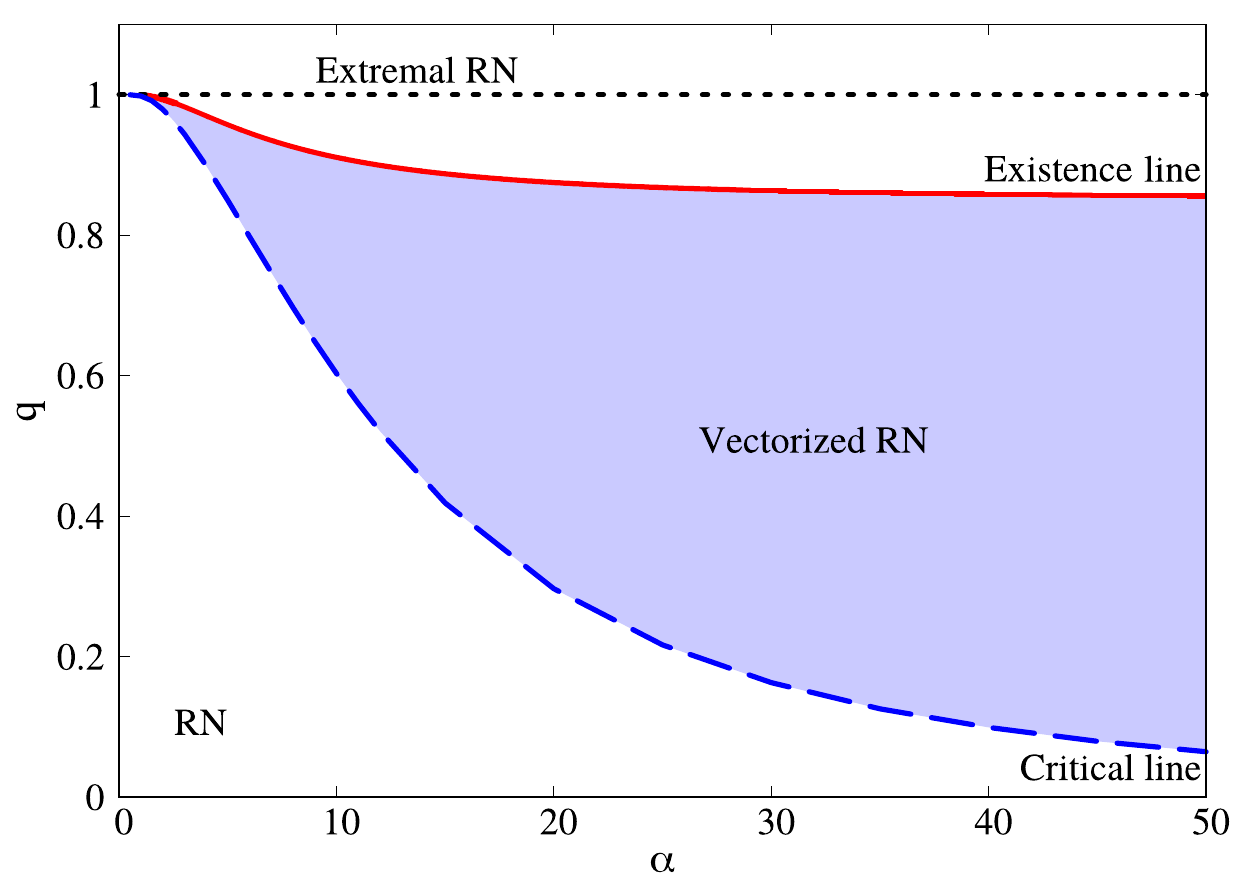}
	 \caption{Graphical representation of the domain of existence.}
	 \label{F3}
	\end{figure}
	Different than the scalar case, all the possible solutions are undercharged, and in fact, the critical line always has a smaller $q$ than the existence line (see Fig.~\ref{F3}). For a fixed $\alpha $ value, one can go from the existence line to the critical line through an increase in $r_H$, meaning that solutions never become singular. Meanwhile, $b_1$ and $\sigma _0$ have a growing increase at the horizon and diverge at the critical line. We have also computed the Kretschmann scalar and observed that the solution is everywhere regular along the domain of existence, including the critical line. 

	Concerning the vector charge, $P$, one observes a monotonic increase along the domain of existence for a fixed $\alpha$. While it starts at zero in the existence line, it grows to almost the double of $Q$ at the critical line.

	Through the study of the domain of existence, one also observes that an increase in $\alpha$ implies a smaller value of the normalized electric charge for both the existence line and the critical line. However, the latter has a faster decrease in $q$ and hence the domain of existence broadens, tending to the Schwarzschild case for $\alpha \rightarrow \infty$ ($q=0.0134$ for $\alpha =100$).

	\medskip

\begin{figure}[h!]
	 \centering
	 \includegraphics[scale=0.66]{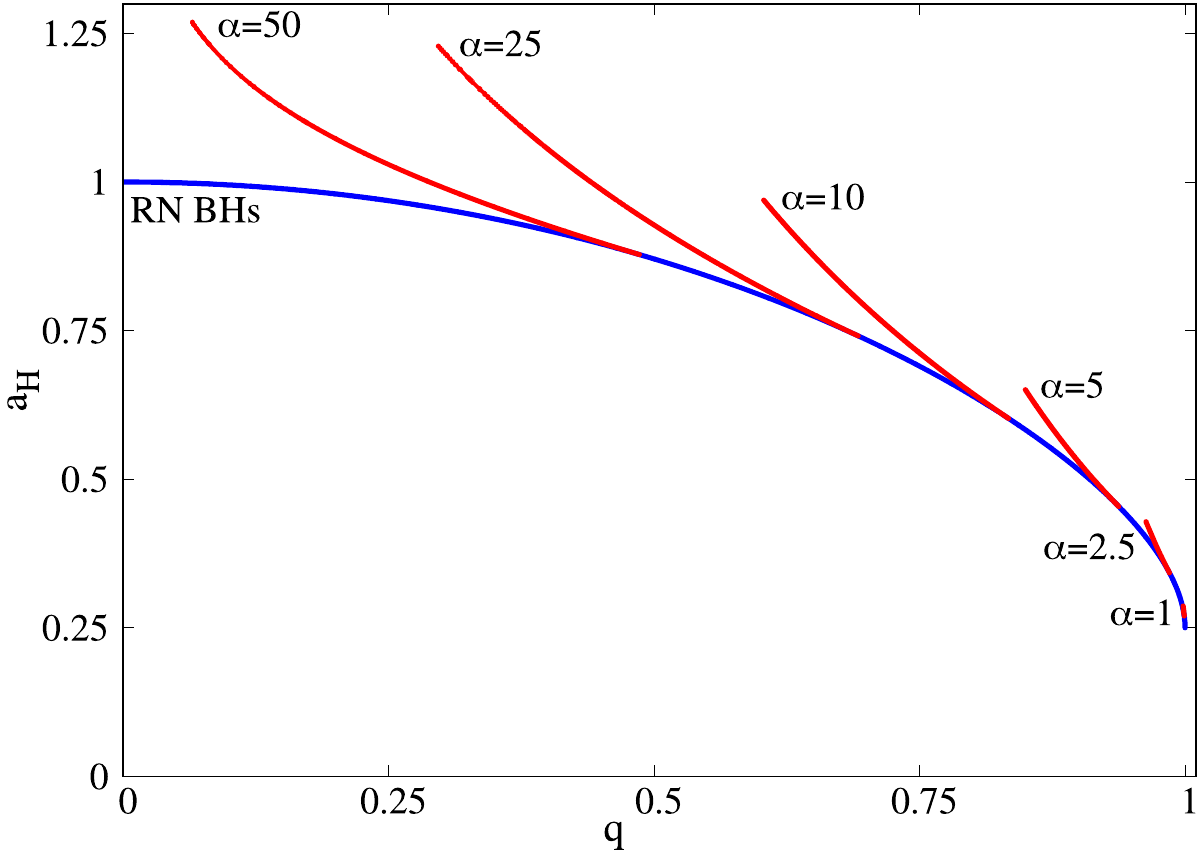}\hfill
	 \includegraphics[scale=0.66]{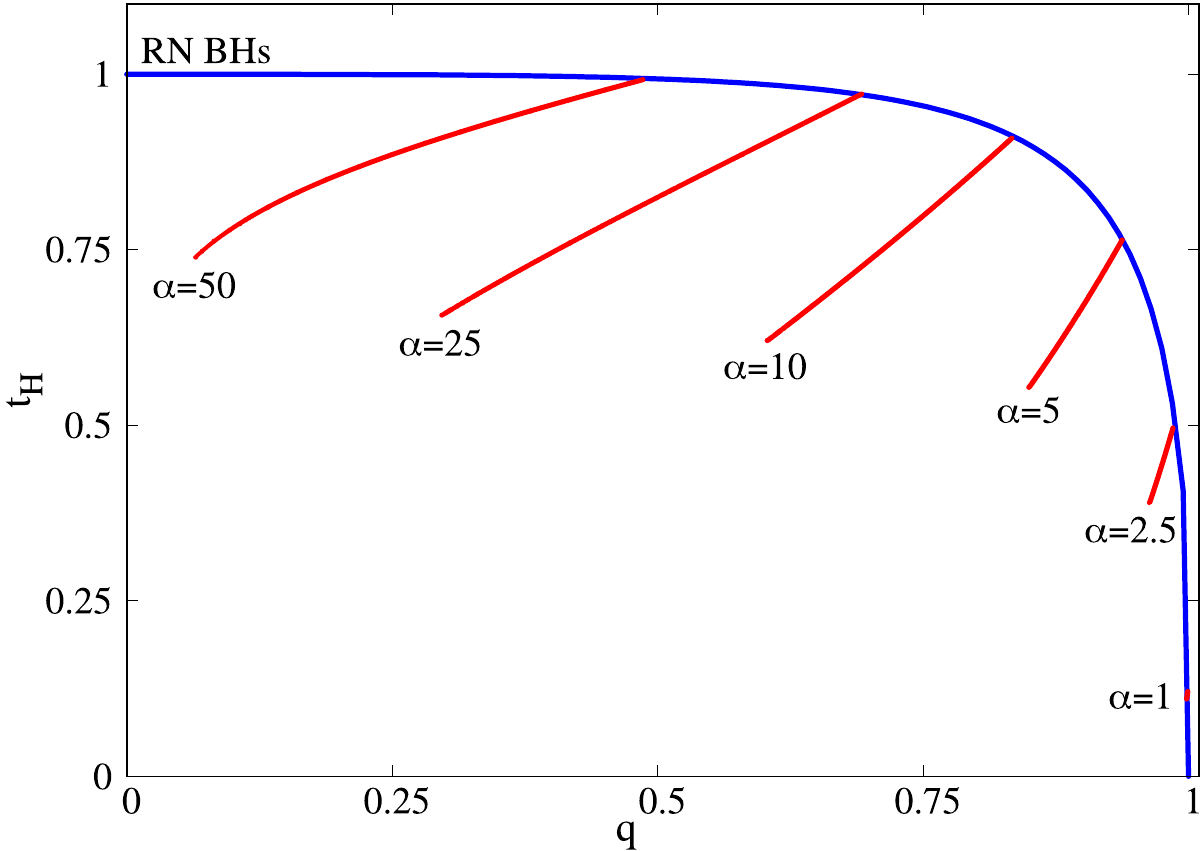}
	 \caption{(Left) Reduced area $a_H$ \textit{vs} reduced charge $q$ and (right) normalized horizon temperature $t_H$ \textit{vs} $q$ for an EMV model. The blue lines represent the non-vectorized RN BH, while the red lines represent the vectorized solutions.}
	 \label{F4}
	\end{figure}

	In addition, to study the thermodynamical preference of vectorized solutions over an equivalent RN BH (see Fig.~\ref{F4}), we have computed the entropy (left) of both solutions and the event horizon temperature (right). 
	
	From the thermodynamical study, we observe an entropic preference of the VRN in relation to an equivalent RN BH, which can be clearly understood by the fact that the ADM Mass of the vectorized BH is smaller than the mass contained in the central BH.
	
	Concerning the horizon temperature (see Fig.~\ref{F4} right), one observes a smaller horizon temperature for the VRN solution for an equivalent RN solution. In addition, the temperature decreases as one goes further from the existence line, however never reaching extremality ($t_H=0$).
	In comparison with the scalarized solutions, both solutions are entropically preferable over an equivalent RN BH. However, while in the scalarized case a solution with a higher coupling constant has a higher entropy, here the opposite occurs. For a SV configuration that allows two coupling constant values, the lower one will be entropically preferable.  	
%
%

\section{Conclusion}\label{S6}
	In this paper, we constructed electrically charged hairy black holes through a spontaneous vectorization process of Reissner-Nordstr\"om Black Holes. To do so, we considered an extra vector field $B_\mu$ that is non-minimally coupled to the Maxwell field.

	Our main objective was to search for stable spherically symmetric configurations. With that in mind, we first showed a no-hair theorem for a generic complex vector field ansatz, along with a flat spacetime no go theorem that was an extension of the first. Notably, this no-hair theorem does not work for a vanishing field frequency $\omega=0$, in which case we have a real vector field with only one component.

	With the last ansatz, we then attempted to construct the vectorized BH solutions. An analytical study of the model showed that, for a non-minimal coupling of the form $f(B^2)=e^{\alpha B^2}$, we have a lower bound for the coupling constant $\alpha \geqslant 1/4$.

	The numerical results show a family of vectorized solutions bifurcating from the RN existence line, reaching a critical solution. Compared to the scalarized RN solutions, the $B_t(r)$ field has a maximum away from the horizon and then slowly decays to zero at infinity. The difference occurs due to the imposition that both at the horizon and infinity, $B_t$ must be zero. 

	A caveat to the vectorized solutions is that they violate the weak energy condition in a small region where the mass function $m(r)$ decreases. Since there is an additional contribution of the interaction term to the energy in the vectorized case, a negative Komar mass outside the horizon emerges, creating a region with negative energy densities.

	Another peculiar property of the vectorized solutions is that they are undercharged, $Q/M<1$. The latter behaviour is amplified by an increasing coupling constant, $\alpha$, tending towards Schwarzschild when $\alpha\rightarrow\infty$. 

	A thermodynamical study tells us that, despite this property, the vectorized solutions are entropically preferred, having a larger horizon area and smaller temperature than the corresponding RN black holes.

    At last, we would like to point out that this is a new generalisation of the EMS model~\cite{Herdeiro:2018wub} and, while most of the behaviour can be extrapolated to the VMS model, there is no guarantee that these solutions are either perturbatively stable or dynamical preferable. Such questions are a research topic for a future and more exhaustive work.

%
\section*{Acknowledgements}
%
We thank comments from Carlos A. R. Herdeiro and Eugen Radu. This work has been supported by Funda\c{c}\~ao para a Ci\^encia e a Tecnologia (FCT), within project UID/MAT/04106/2019 (CIDMA), by CENTRA (FCT) strategic project UID/FIS/00099/2020, by national funds (OE), through FCT, I.P., in the scope of the framework contract foreseen in the numbers 4, 5 and 6 of the article 23, of the Decree-Law 57/2016, of August 29, changed by Law 57/2017, of July 19. J. M. S. Oliveira  is supported by an FCT post-doctoral grant through the project PTDC/FIS-OUT/28407/2017 and A. M. Pombo is supported by the FCT grant PD/BD/142842/2018. This work has further been supported by  the  European  Union's  Horizon 2020 research and innovation (RISE) programmes H2020-MSCA-RISE-2015 Grant No.~StronGrHEP-690904 and H2020-MSCA-RISE-2017 Grant No.~FunFiCO-777740. The authors would like to acknowledge networking support by the COST Action CA16104. At last, we acknowledge support from the  projects CERN/FIS-PAR/0027/2019  and  PTDC/FIS-AST/3041/2020.


  \bibliographystyle{ieeetr}
  \bibliography{biblio}

\begin{thebibliography}{10}

\bibitem{Doneva:2017bvd}
D.~D. Doneva and S.~S. Yazadjiev, ``{New Gauss-Bonnet Black Holes with
  Curvature-Induced Scalarization in Extended Scalar-Tensor Theories},'' {\em
  Phys. Rev. Lett.}, vol.~120, no.~13, p.~131103, 2018.

\bibitem{Silva:2017uqg}
H.~O. Silva, J.~Sakstein, L.~Gualtieri, T.~P. Sotiriou, and E.~Berti,
  ``{Spontaneous scalarization of black holes and compact stars from a
  Gauss-Bonnet coupling},'' {\em Phys. Rev. Lett.}, vol.~120, no.~13,
  p.~131104, 2018.

\bibitem{Antoniou:2017acq}
G.~Antoniou, A.~Bakopoulos, and P.~Kanti, ``{Evasion of No-Hair Theorems and
  Novel Black-Hole Solutions in Gauss-Bonnet Theories},'' {\em Phys. Rev.
  Lett.}, vol.~120, no.~13, p.~131102, 2018.

\bibitem{Damour:1993hw}
T.~Damour and G.~Esposito-Farese, ``{Nonperturbative strong field effects in
  tensor - scalar theories of gravitation},'' {\em Phys. Rev. Lett.}, vol.~70,
  pp.~2220--2223, 1993.

\bibitem{Herdeiro:2018wub}
C.~A. Herdeiro, E.~Radu, N.~Sanchis-Gual, and J.~A. Font, ``{Spontaneous
  Scalarization of Charged Black Holes},'' {\em Phys. Rev. Lett.}, vol.~121,
  no.~10, p.~101102, 2018.

\bibitem{Cunha:2019dwb}
P.~V. Cunha, C.~A. Herdeiro, and E.~Radu, ``{Spontaneously Scalarized Kerr
  Black Holes in Extended Scalar-Tensor\textendash{}Gauss-Bonnet Gravity},''
  {\em Phys. Rev. Lett.}, vol.~123, no.~1, p.~011101, 2019.

\bibitem{Herdeiro:2020wei}
C.~A. Herdeiro, E.~Radu, H.~O. Silva, T.~P. Sotiriou, and N.~Yunes,
  ``{Spin-induced scalarized black holes},'' 9 2020.

\bibitem{Berti:2020kgk}
E.~Berti, L.~G. Collodel, B.~Kleihaus, and J.~Kunz, ``{Spin-induced black-hole
  scalarization in Einstein-scalar-Gauss-Bonnet theory},'' 9 2020.

\bibitem{dima2020spin}
A.~Dima, E.~Barausse, N.~Franchini, and T.~P. Sotiriou, ``Spin-induced black
  hole spontaneous scalarization,'' {\em arXiv preprint arXiv:2006.03095},
  2020.

\bibitem{fernandes2019spontaneous}
P.~G. Fernandes, C.~A. Herdeiro, A.~M. Pombo, E.~Radu, and N.~Sanchis-Gual,
  ``Spontaneous scalarisation of charged black holes: coupling dependence and
  dynamical features,'' {\em Classical and Quantum Gravity}, vol.~36, no.~13,
  p.~134002, 2019.

\bibitem{astefanesei2019einstein}
D.~Astefanesei, C.~Herdeiro, A.~Pombo, and E.~Radu, ``Einstein-maxwell-scalar
  black holes: classes of solutions, dyons and extremality,'' {\em Journal of
  High Energy Physics}, vol.~2019, no.~10, p.~78, 2019.

\bibitem{fernandes2019charged}
P.~G. Fernandes, C.~A. Herdeiro, A.~M. Pombo, E.~Radu, and N.~Sanchis-Gual,
  ``Charged black holes with axionic-type couplings: Classes of solutions and
  dynamical scalarization,'' {\em Physical Review D}, vol.~100, no.~8,
  p.~084045, 2019.

\bibitem{blazquez2020einstein}
J.~L. Bl{\'a}zquez-Salcedo, C.~A. Herdeiro, J.~Kunz, A.~M. Pombo, and E.~Radu,
  ``Einstein-maxwell-scalar black holes: the hot, the cold and the bald,'' {\em
  Physics Letters B}, p.~135493, 2020.

\bibitem{blazquez2020quasinormal}
J.~L. Bl{\'a}zquez-Salcedo, C.~A. Herdeiro, S.~Kahlen, J.~Kunz, A.~M. Pombo,
  and E.~Radu, ``Quasinormal modes of hot, cold and bald
  einstein-maxwell-scalar black holes,'' {\em arXiv preprint arXiv:2008.11744},
  2020.

\bibitem{Astefanesei:2020qxk}
D.~Astefanesei, C.~Herdeiro, J.~a. Oliveira, and E.~Radu, ``{Higher dimensional
  black hole scalarization},'' {\em JHEP}, vol.~09, p.~186, 2020.

\bibitem{baum1970vector}
P.~F. Baum, ``Vector fields and gauss-bonnet,'' {\em Bulletin of the American
  Mathematical Society}, vol.~76, no.~6, pp.~1202--1211, 1970.

\bibitem{gleiser2005linear}
R.~J. Gleiser and G.~Dotti, ``Linear stability of einstein-gauss-bonnet static
  spacetimes: Vector and scalar perturbations,'' {\em Physical Review D},
  vol.~72, no.~12, p.~124002, 2005.

\bibitem{herdeiro2016kerr}
C.~Herdeiro, E.~Radu, and H.~Runarsson, ``Kerr black holes with proca hair,''
  {\em Classical and Quantum Gravity}, vol.~33, no.~15, p.~154001, 2016.

\bibitem{fan2016black}
Z.-Y. Fan, ``Black holes with vector hair,'' {\em Journal of High Energy
  Physics}, vol.~2016, no.~9, p.~39, 2016.

\bibitem{annulli2019electromagnetism}
L.~Annulli, V.~Cardoso, and L.~Gualtieri, ``Electromagnetism and hidden vector
  fields in modified gravity theories: Spontaneous and induced vectorization,''
  {\em Physical Review D}, vol.~99, no.~4, p.~044038, 2019.

\bibitem{ramazanouglu2017spontaneous}
F.~M. Ramazano{\u{g}}lu, ``Spontaneous growth of vector fields in gravity,''
  {\em Physical Review D}, vol.~96, no.~6, p.~064009, 2017.

\bibitem{Ramazanoglu:2019gbz}
F.~M. Ramazano\u{g}lu, ``{Spontaneous tensorization from curvature coupling and
  beyond},'' {\em Phys. Rev. D}, vol.~99, no.~8, p.~084015, 2019.

\bibitem{minamitsuji2020spontaneous}
M.~Minamitsuji, ``Spontaneous vectorization in the presence of vector-field
  coupling to matter,'' {\em Physical Review D}, vol.~101, no.~10, p.~104044,
  2020.

\bibitem{Kase:2020yhw}
R.~Kase, M.~Minamitsuji, and S.~Tsujikawa, ``{Neutron stars with a generalized
  Proca hair and spontaneous vectorization},'' {\em Phys. Rev. D}, vol.~102,
  no.~2, p.~024067, 2020.

\bibitem{Ramazanoglu:2019jrr}
F.~M. Ramazano\u{g}lu and K.~I. \"Unl\"ut\"urk, ``{Generalized disformal
  coupling leads to spontaneous tensorization},'' {\em Phys. Rev. D}, vol.~100,
  no.~8, p.~084026, 2019.

\bibitem{misner1964relativistic}
C.~W. Misner and D.~H. Sharp, ``Relativistic equations for adiabatic,
  spherically symmetric gravitational collapse,'' {\em Physical Review},
  vol.~136, no.~2B, p.~B571, 1964.

\bibitem{Pena:1997cy}
I.~Pena and D.~Sudarsky, ``{Do collapsed boson stars result in new types of
  black holes?},'' {\em Class. Quant. Grav.}, vol.~14, pp.~3131--3134, 1997.

\bibitem{Herdeiro:2019iwl}
C.~A. Herdeiro, J.~a.~M. Oliveira, and E.~Radu, ``{A class of solitons in
  Maxwell-scalar and Einstein\textendash{}Maxwell-scalar models},'' {\em Eur.
  Phys. J. C}, vol.~80, no.~1, p.~23, 2020.

\bibitem{derrick1964comments}
G.~Derrick, ``Comments on nonlinear wave equations as models for elementary
  particles,'' {\em Journal of Mathematical Physics}, vol.~5, no.~9,
  pp.~1252--1254, 1964.

\bibitem{cunha2017light}
P.~V. Cunha, E.~Berti, and C.~A. Herdeiro, ``Light-ring stability for
  ultracompact objects,'' {\em Physical Review Letters}, vol.~119, no.~25,
  p.~251102, 2017.

\bibitem{cunha2020stationary}
P.~V. Cunha and C.~A. Herdeiro, ``Stationary black holes and light rings,''
  {\em Physical Review Letters}, vol.~124, no.~18, p.~181101, 2020.

\end{thebibliography}


\end{document}